\documentclass{aastex63}

\usepackage{graphicx,epsfig}
\usepackage{dcolumn}
\usepackage{bm}
\usepackage{xcolor}
\usepackage{multibib}
\newcites{main}{References}
\newcites{appendix}{References (Appendix)}
\usepackage{subfigure}
\usepackage{soul}
\usepackage{tcolorbox}
\usepackage{amsmath}
\usepackage{verbatim}

\newcommand{\be}{\begin{equation}}
\newcommand{\ee}{\end{equation}}
\newcommand{\bse}{\begin{subequations}}
\newcommand{\ese}{\end{subequations}}
\newcommand{\bary}{\begin{eqnarray}}
\newcommand{\eary}{\end{eqnarray}}
\newcommand{\bwt}{\begin{widetext}}
\newcommand{\ewt}{\end{widetext}}

\begin{document}

\title{A two-zone photohadronic scenario for EHBL-like behavior of Mrk 501}


\author{Sarira Sahu}
\email{sarira@nucleares.unam.mx}
\affiliation{Instituto de Ciencias Nucleares, Universidad Nacional Aut\'onoma de M\'exico, \\
Circuito Exterior, C.U., A. Postal 70-543, 04510 Mexico DF, Mexico}

\author{Carlos E. L\'opez Fort\'in}
\email{carlos.fortin@correo.nucleares.unam.mx}
\affiliation{Instituto de Ciencias Nucleares, Universidad Nacional Aut\'onoma de M\'exico, \\
Circuito Exterior, C.U., A. Postal 70-543, 04510 Mexico DF, Mexico}

\author{Luis H. Castañeda Hernández}
\email{pabcdarioluis@hotmail.com}
\affiliation{Facultad de Ciencias, Universidad Nacional Aut\'onoma de M\'exico, \\
Circuito Exterior, C.U., A. Postal 70-543, 04510 Mexico DF, Mexico}

\author{Shigehiro Nagataki}
\email{shigehiro.nagataki@riken.jp}
\affiliation{Astrophysical Big Bang Laboratory, RIKEN,\\
Hirosawa, Wako, Saitama 351-0198, Japan} 
\affiliation{Interdisciplinary Theoretical \& Mathematical Science (iTHEMS),\\
RIKEN, Hirosawa, Wako, Saitama 351-0198, Japan}

\author{Subhash Rajpoot}
\email{Subhash.Rajpoot@csulb.edu}
\affiliation{Department of Physics and Astronomy, California State University,\\ 
1250 Bellflower Boulevard, Long Beach, CA 90840, USA}



\begin{abstract}
Major outbursts have been observed from the well-known high-energy peaked blazar Markarian 501 since its discovery in 1996. Especially, two episodes of very high energy gamma-ray flaring events during May-July 2005 and June 2012 are of special significance, when the source exhibited extreme HBL-like behavior. The successful standard photohadronic model seems inadequate to explain these extraneous behaviors. We propose a two-zone photohadronic scenario to overcome this problem. In this picture, the low energy regime (zone-1) of the spectrum follows the standard photohadronic interpretation, while the high energy regime (zone-2) of the spectrum is new, with a spectral index $\delta_2\geq 3.1$, which is solely due to the extreme nature of the flaring event. We also estimate the bulk Lorentz factor corresponding to these extreme flaring events. By analyzing many flaring events before and after these extreme events we argue that the extreme HBL-like events are transient and may repeat in future.
\end{abstract}


\section{Introduction}

Blazars are a subclass of active galactic nuclei (AGNs) exhibiting very rapid flux variability in the entire electromagnetic spectrum from radio to very-high energy (VHE, above 100 GeV) $\gamma$-rays \citep{Acciari:2010aa}.
The electromagnetic spectrum is produced in a highly relativistic jet pointing towards the observer's line of sight \citep{Urry:1995mg}. The small viewing angle is responsible for the strong relativistic effects, e.g. boosting of the emitted power and shortening of the time scale \citep{Abdo:2010rw}.

Blazars have spectral energy distribution (SED) characterized by two non-thermal peaks \citep{Dermer:1993cz}. The first peak, between infrared to X-ray energy is produced by the synchrotron emission from a population of relativistic electrons in the jet.
The second peak, in X-rays to $\gamma$-ray, is believed to be produced either from the Synchrotron Self-Compton (SSC) scattering of high-energy electrons with the low-energy self-produced synchrotron photons in the jet \citep{Maraschi:1992iz,Murase:2011cy,Gao:2012sq} or from the external Compton scattering (EC) with external sources such as photons from the accretion disk or broad-line regions (leptonic models) \citep{Sikora:1994zb,Blazejowski:2000ck}. 

BL Lac objects can be further classified depending on the frequency of the synchrotron peak as: low-energy peaked blazars (LBLs, $\nu_{peak} < 10^{14}\, Hz$), intermediate-energy peaked blazars (IBLs, $\nu_{peak}$ between $10^{14}\, Hz$ and $10^{15}\, Hz$), and high energy-peaked blazars (HBLs, $\nu_{peak}$ between $10^{15}\, Hz$ and $10^{17}\, Hz$) \citep{Abdo:2009iq}. A new subclass is proposed \citep{Costamante:2001pu} with extreme spectral properties and energy shifted towards the edge of the blazar sequence. This is named extreme high energy peaked blazar (EHBL). Its synchrotron peak is shifted towards the higher energies ($ >10^{17}\, Hz$) with respect to conventional HBLs. EHBLs are characterized by low luminosity and limited variability and are difficult to be detected by the current gamma-rays surveys.
Up to now, only a few EHBLs have been detected in TeV energy and some well known objects are 1ES 0229+200, 1ES 0347-232, RGB J0710+591 and 1ES 1101-232 \citep{Costamante:2017xqg}. Among all, 1ES 0229+200 has the highest high energy peak frequency. Apart from the above EHBLs, two nearby HBLs, Markarian 501 (Mrk 501) and 1ES 1959+650 have also EHBL-like behavior with harder TeV spectra and shifted synchrotron peaks during some flaring episodes \citep{Ahnen:2018mtr,Hayashida:2020wez}. 
Also, as discussed by Faffano et al. \citep{Foffano:2019itc} the EHBL class might be a complex population of sources, characterized by different spectral properties and these different behaviors at VHE gamma-rays might be characterizing different sub-classes within the EHBL class.

Theoretically it is challenging to explain the shift of the second peak towards higher energy and with a hard spectrum in terms of the standard one-zone leptonic SSC model as it predicts softer SSC spectrum in the Klein-Nishina regime, contrary to  observation \citep{Paggi:2009yx}.
However, these enigmatic spectra can still be explained using the leptonic model at a price of introducing unrealistically large model parameters, such as the values of minimum electron Lorentz factor and bulk Lorentz factor \citep{Ahnen:2018mtr,Hayashida:2020wez}. It also needs a very low magnetic field. 
Several other solutions are also proposed, such as: two-zone leptonic model, IC scattering of the electron with the cosmic microwave background, spine-layer structured jet model and variants of hadronic model \citep{Boettcher:2008fh,Acciari:2019ntl,Ahnen:2018mtr}.

VHE $\gamma$-rays from astrophysical sources are known to undergo energy-dependent attenuation by interacting with the extragalactic background light (EBL) via electron-positron pair production \citep{Ackermann:2012sza}. This attenuation affects the shape of the spectrum at very high energies and several models exist to account for this attenuation at different redshifts \citep{Franceschini:2008tp,Dominguez:2010bv}. 
As EHBLs are an emerging class of BL Lac objects with extreme spectral properties, particularly in the TeV range, they are an ideal probe for EBL \citep{Tavecchio:2013fwa,Tavecchio:2015cid}. Moreover, the association of IceCube neutrino event(s) with blazar(s) \citep{Padovani:2016wwn, IceCube:2018dnn, IceCube:2018cha} makes these sources even more interesting to study in further detail.  

Previously, the photohadronic model has been used to explain very well the multi-TeV flaring from many HBLs including Mrk 501 \citep{Sahu:2019kfd,Sahu:2019scf}. This well studied Mrk 501 has shown EHBL-like behavior, during the VHE flaring in 2005 and 2012 \citep{Albert:2007zd,Ahnen:2018mtr}. So, it is natural to extend the photohadronic scenario to explain the extraneous behavior of these flaring events. However, the photohadronic model in its standard form with EBL correction does not explain the VHE spectra, although most of the previous flarings were very well explained  \citep{Sahu:2019kfd}. Here, for the first time, we demonstrated that, EHBL-like behavior of Mrk 501 can be explained by the two-zone photohadronic model. The low energy part of the spectrum is the standard HBL flaring event, and the high energy part is due to the extreme nature of the event. We have also argued that, these EHBL-like events are transient in nature and may repeat in the future.

\section{Photohadronic Model}

The multi-TeV flaring from the HBLs are explained using the photohadronic model \citep{Sahu:2019lwj,Sahu:2019kfd}.
In this model, the Fermi-accelerated protons in the jet interact with the low-energy background seed photons through $p\gamma\rightarrow\Delta^+$ process, which subsequently decays into $\gamma$-rays through intermediate $\pi^0$ and to neutrinos via $\pi^+$. The neutrinos produced during the flaring period have energies about half of the photon energy \citep{Sahu:2019lwj} and are not interesting from the IceCube point of view as the energy is very low. The photohadronic 
scenario discussed here is based on the standard interpretation of the leptonic model, where the low and high energy peaks have leptonic origin 
and the emitting region is a blob of comoving radius $R'_b$ (where $'$ implies comoving frame), moving with a bulk Lorentz factor $\Gamma$ and Doppler factor $\mathcal{D}$. The injected proton spectrum is a power-law in its energy $E_p$ \citep{Dermer:1993cz} as
$dN/dE_p\propto E^{-\alpha}_p$,
where the spectral index $\alpha \ge 2$. In order to produce the $\Delta$-resonance, the $E_p$ and the seed photon energy $\epsilon_\gamma$ must satisfy the kinematical condition \citep{Sahu:2019lwj},
\be
E_p \epsilon_\gamma=\frac{0.32\ \Gamma\mathcal{D}}{(1+z)^{2}}\ \mathrm{GeV^2}.
\label{eq:kinproton}
\ee
For HBLs, $\Gamma \approx \mathcal{D}$ and $z$ is the redshift of the object. The observed VHE $\gamma$-ray has energy $E_{\gamma}=0.1\, E_p$. 

In a canonical jet scenario the seed photon density is low, thus the efficiency for the $\Delta$-resonance process is low. So super-Eddington power in proton is required to account for the observed VHE flux \citep{Cao:2014nia}. To circumvent this problem, and based on evidence of complex structure of AGN jets during the multi-TeV flaring phase \citep{Ghisellini:2004ec}, the photohadronic scenario assumes a double jet structure where an inner jet of size $R'_f$ and photon density $n'_{\gamma,f}$ is surrounded by an outer jet of size $R'_b$ and photon density $n'_{\gamma}$, with $R'_f<R'_b$ and $n'_{\gamma,f}>n'_{\gamma}$.
The geometry of the double-jet scenario is shown in Figure 1 of \citep{Sahu:2019lwj}.
Due to the adiabatic expansion of the jet, the photon density of inner jet decreases as it crosses into the outer jet. As the inner photon density is unknown, 
a scaling behavior is assumed \citep{Sahu:2019kfd}. This leads to the relation,
\be
\frac{n'_{\gamma,f}(\epsilon_{\gamma,1})}{n'_{\gamma,f}(\epsilon_{\gamma,2})} \simeq\frac{n'_{\gamma}(\epsilon_{\gamma,1})}{n'_{\gamma}(\epsilon_{\gamma,2})}.
\label{eq:scaling}
\ee
It shows that, during multi-TeV flaring, the ratios of the photon densities at energies $\epsilon_{\gamma,1}$ and  $\epsilon_{\gamma,2}$ in the inner jet and the outer jet region are almost the same. The photon density in the outer region can be calculated from the observed flux and using Eq. (\ref{eq:scaling}), we can express the inner photon density in terms of the observed flux.
It is observed that, the range of $E_{\gamma}$ corresponds to  the seed photon energy  $\epsilon_{\gamma}$ in the low energy tail region of the SSC spectrum and in this region the SSC flux is a perfect power-law, given as $\Phi_{SSC}\propto \epsilon^{\beta}_{\gamma}$ \citep{Sahu:2019lwj}.
Moreover, the observed VHE $\gamma$-ray flux is proportional to the high energy incident proton flux $F_p$ and the seed photon density in the jet background, $n'_{\gamma,f}$. Using the relation in Eq. (\ref{eq:scaling}) and taking into account the EBL correction \citep{Franceschini:2008tp}, the observed multi-TeV spectrum is expressed in terms of the intrinsic flux $F_{\gamma, in}$ as 
\be
F_{\gamma,obs}(E_{\gamma})=F_0 \left ( \frac{E_\gamma}{TeV} \right )^{-\delta+3}\,e^{-\tau_{\gamma\gamma}(E_\gamma,z)}=F_{\gamma, in}(E_{\gamma})\, e^{-\tau_{\gamma\gamma}(E_\gamma,z)},
\label{eq:fluxgeneral}
\ee
where $F_0$ is the normalization constant which can be fixed from the observed VHE spectrum and $\delta=\alpha+\beta$ is the only free parameter in this model. $F_{\gamma,in}$ is the intrinsic VHE gamma-ray flux. Previous studies have shown that $\delta$ lies in the range $2.5\le \delta \le 3.0$. Moreover, flaring states can be roughly classified into three categories, depending on the value of $\delta$ as: (i) very high state with $2.5\le\delta\le 2.6$, (ii) high state with $2.6<\delta<3.0$ and (iii) low state for $\delta=3.0$ \citep{Sahu:2019kfd}.

As the proton spectral index is $\alpha\ge 2.0$, this automatically constrains the value of $\beta$ for a given $\delta$. But, here we take $\alpha=2$ which is the generally accepted value. The value of $\beta$ can be obtained independently, by fitting the low energy tail region of the SSC SED from a simultaneous observation along with the VHE spectrum. However, simultaneous observations of the SSC SED in the tail region during the VHE flaring are rare and most of them are either non-simultaneous or fitted using the leptonic models. The value of $\delta$ allows one to characterize the flaring state for a given observation without depending explicitly on the modeling of the SSC region.

\section{Flaring history of Mrk 501}

Mrk 501 (z=0.034) is the second well known bright BL Lac object of HBL type and one of the brightest extragalactic sources in the X-ray/TeV sky. It was first detected in VHE by Whipple telescopes in 1996 \citep{Abdo:2009wu} and since then several major outburst in multi-TeV have been observed and it has been the target of many multiwavelength campaigns
mainly covering VHE flaring activities
\citep{Kataoka:1998mw,Albert:2007me,Kranich:2009ht,Aleksic:2014usa} 
The outburst of 1997, exhibited an unprecedented flare in VHE gamma-rays with an integral flux of up to four times the magnitude of the Crab Nebula (CU) flux \citep{Petry:2000pb}.
As part of multiwavelength campaign 
from March 15 to August 1, 2009, it was observed by $\sim 30$ different instruments covering the entire electromagnetic spectrum \citep{Aliu:2016kzx,Ahnen:2016hsf}.

From May to June 2005, Mrk 501 was observed for 30 nights with an overall observation time of 54.8 hr and flaring above 0.10 TeV was observed by MAGIC telescopes with an order of magnitude flux variation \citep{Albert:2007zd}. On the nights of June 30 and July 9, flux-doubling was observed within a time span of 2 minutes and this is the fastest flux variation ever observed from Mrk 501. During this period, the synchrotron peak as well as the high energy peak were also shifted towards higher energy limits which is consistent with the EHBL interpretation of Mrk 501.

Similarly during a multiwavelength campaign between March and July of 2012, more than 25 instruments including MAGIC, First G-APD Cherenkov Telescope (FACT) and VERITAS were employed when flux above 0.2 TeV energy was observed \citep{Ahnen:2018mtr}. The highest activity occurred on June 9 with the peak flux of 0.3 CU around 0.2 TeV. It was observed that the spectral indices of both the X-ray and VHE gamma-ray were extremely hard and this is the hardest VHE spectra measured from Mrk 501 to date. Throughout this observation period, the synchrotron peak shifted above 5 keV and the SSC peak above 0.5 TeV, thus behaving like EHBL. Apart from the above active emission states, few more VHE flaring events were also observed \citep{Chandra:2017vkw,Abdalla:2019krx}.

\begin{figure}[h]
{\centering
\resizebox*{0.8\textwidth}{0.5\textheight}
{\includegraphics{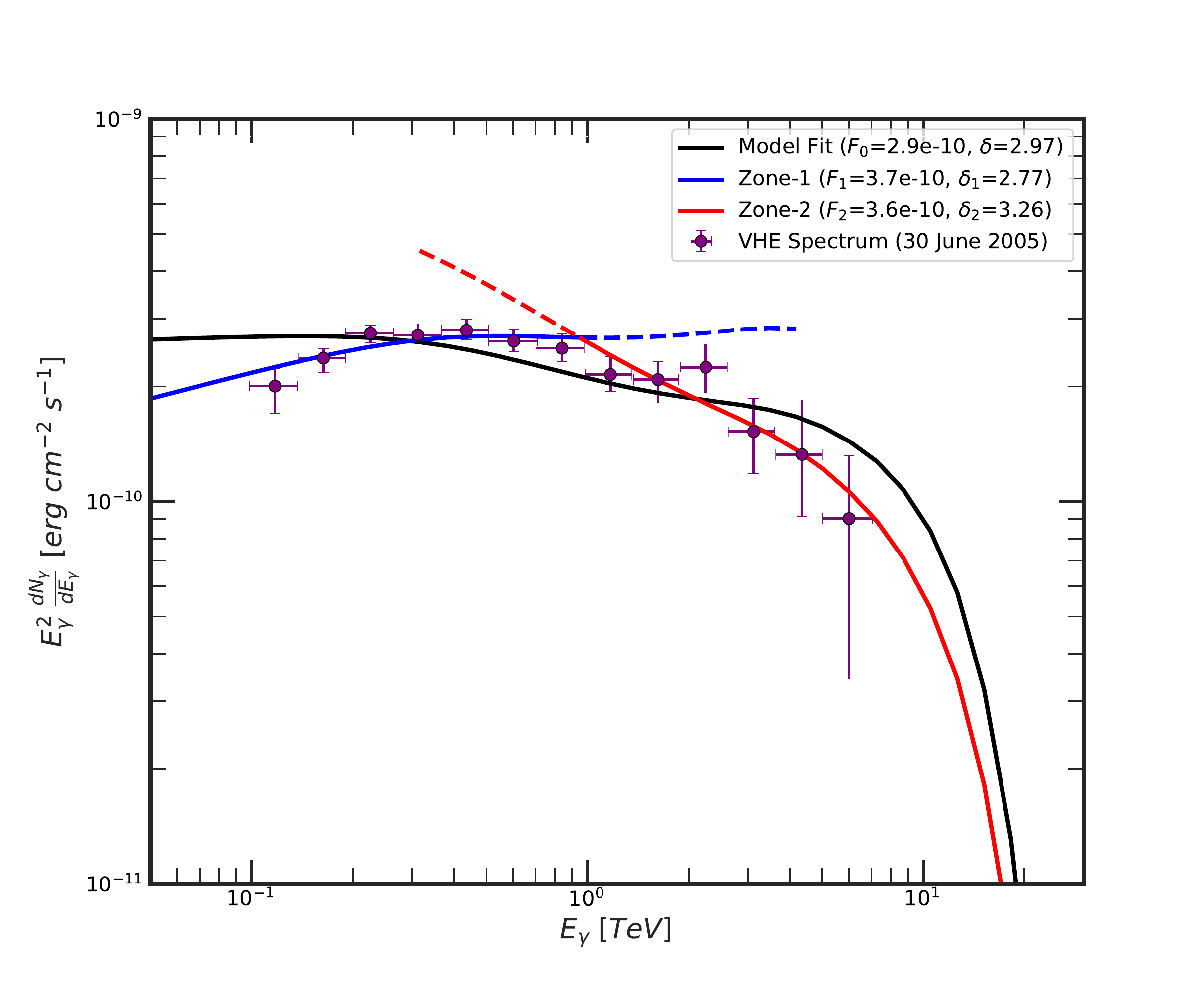}
}
\par}
\caption{
The VHE spectrum of Mrk 501 observed in the night of 30th June 2005 by MAGIC telescopes is fitted with the photohadronic model (black). In all the figures the normalization constants $F_i$ ($i=0,1,2$) are defined in units of $erg\, cm^{-2} \, s^{-1}$. The best fit to Zone-1 is shown in the blue curve
with a statistical goodness of 99.9\%. The dashed blue curve shows the behavior of the model in the high energy limit. Similarly, the spectrum in Zone-2 is fitted with the red curve with a statistical significance of 99.7\% and the dashed red curve shows its behavior in the low energy limit. In Figures \ref{fig:figure2} and \ref{fig:figure3}, the dashed curves have the same interpretation as this figure.
}
\label{fig:figure1}
\end{figure}
\begin{figure}[h]
{\centering
\resizebox*{0.8\textwidth}{0.5\textheight}
{\includegraphics{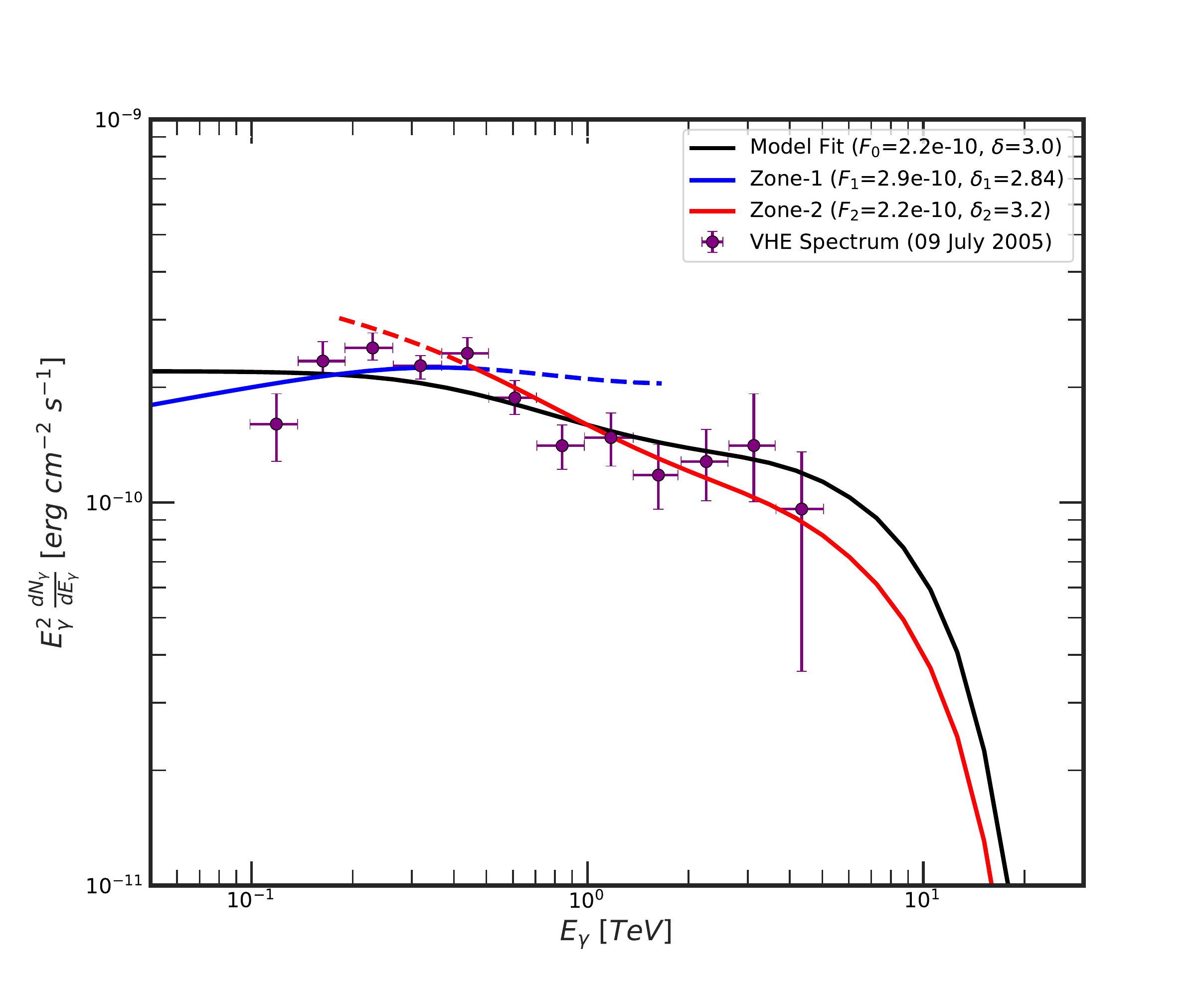}}
\par}
\caption{
The VHE spectrum of Mrk 501 observed in the night of 9th July 2005. Different curves have the same interpretation as in Figure \ref{fig:figure1}.
}
\label{fig:figure2}
\end{figure}

\begin{figure}[h]
{\centering
\resizebox*{0.8\textwidth}{0.5\textheight}
{\includegraphics{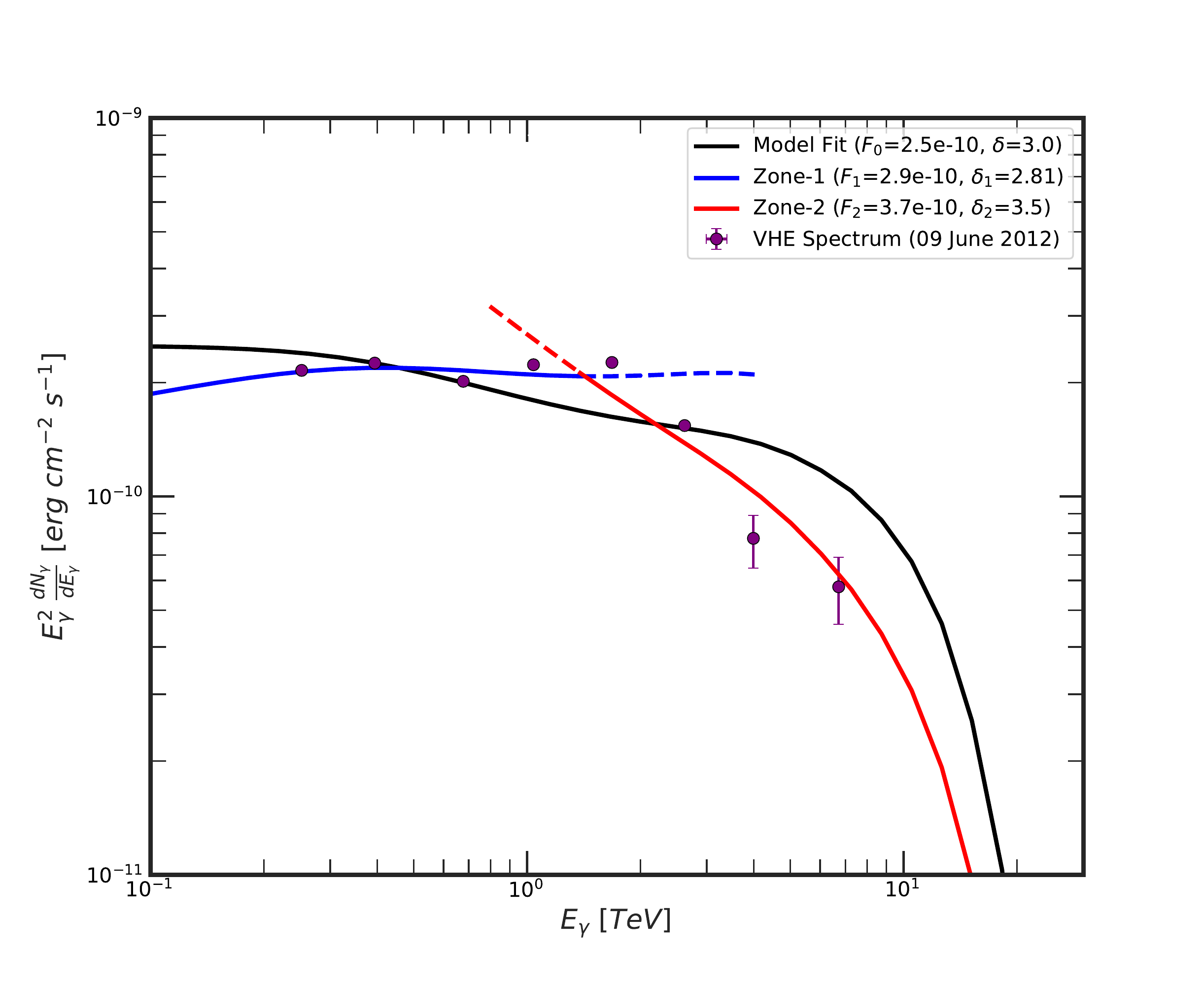}}
\par}
\caption{
The VHE spectrum of Mrk 501 observed in the night of June 9, 2012. Different curves have the same interpretation as Figure \ref{fig:figure1}.
}
\label{fig:figure3}
\end{figure}


\subsection{Multi-TeV EHBL-like events}

As discussed above, Mrk 501 was observed by MAGIC telescopes during May-July 2005 and an order of magnitude variation in VHE flux was observed \citep{Albert:2007zd}. 
Marked enhancement in flux was observed on June 30 ($3.48\pm 0.10$ CU) and July 9 ($3.12\pm 0.12$ CU), when the source was in a very active state. During these two nights the synchrotron peaks were above $10^{17}$ Hz and the second peak position in both cases was also shifted towards higher energy. As the photohadronic model is very successful in explaining the multi-TeV spectra of HBLs, by default, first we use it to fit these spectra before trying other options and the best fits are shown in black curves in Figure \ref{fig:figure1}  and \ref{fig:figure2}. However, our fits are not good. Also, the observed spectra falls faster than the fitted curves and in low energy regime the behavior of the data and the curves is very different. We also fitted the VHE spectrum of June 9, 2012 \citep{Ahnen:2018mtr} which is shown in black curve in Figure \ref{fig:figure3}. It is clearly seen that the fit is poor. Also the observed spectrum falls faster than the fitted curve. From above examples it is obvious that the standard photohadronic model is inadequate to explain the observed spectra. 

In the photohadronic scenario, the high energy regime of the VHE spectrum is produced from the higher energy part of the proton spectrum in the jet, and the low energy regime of the VHE spectrum is from the lower part of the proton spectrum, both satisfying the kinematical condition in Eq.(\ref{eq:kinproton}). As $E_{\gamma} \propto \epsilon^{-1}_{\gamma}$, in EHBL flaring, the SED is shifted towards higher energy as compared to HBL flaring. As a result, photohadronic model fails to explain the VHE spectrum simply because the seed photon flux is no more a single power-law. 

Observationally, the transition from HBL to EHBL-like is the result of lateral displacement of both the peaks in the SED towards higher energies. As a result, jet parameters, such as bulk Lorentz factor, blob size, magnetic field etc may be different from HBL, but we assume that the mechanisms of particle acceleration and emission should remain the same. Therefore, the photohadronic process should still be the dominant one. We have already shown that, a single power-law is inadequate to fit the observed spectrum, so, we assume two power-laws for the background seed photon flux in the SSC band, expressed as
\be
\Phi_{SSC}\propto
 \left\{ 
\begin{array}{cr}
E^{-\beta_1}_{\gamma}
, & \quad 
 \mathrm{100\, GeV\, \lesssim E_{\gamma} \lesssim E^{intd}_{\gamma}}
\\ E^{-\beta_2}_{\gamma} ,
& \quad   \mathrm{E_{\gamma}\gtrsim E^{intd}_{\gamma}}
\\
\end{array} \right. .
\label{eq:sscflux}
\ee
The spectral indices $\beta_1$ and $\beta_2$ are not equal ($\beta_1\neq \beta_2$). $E^{intd}_{\gamma}$ is an energy scale around which the transition between zone-1 and zone-2 takes place and its value can be fixed from the individual flaring spectrum. Using this, the observed spectrum can be expressed as
\be
F_{\gamma, obs}=
e^{-\tau_{\gamma\gamma}}\times
\begin{cases}
 F_1 \, \left ( \frac{E_{\gamma}}{TeV} \right )^{-\delta_1+3}
, & \quad 
\mathrm{100\, GeV\, \lesssim E_{\gamma} \lesssim E^{intd}_{\gamma}}\,\,\,\,\, (\text{zone-1}) \\ 
F_2 \, \left ( \frac{E_{\gamma}}{TeV} \right )^{-\delta_2+3},
& \quad \,\,\,\,\,\,\,\,\,\,\,\,\,\,\,\,\,\,\,\,\,\,\,\,\,\,\,\,\,\,\, \mathrm{E_{\gamma}\gtrsim E^{intd}_{\gamma}}\,\,\,\,\, (\text{zone-2})
 \end{cases},
\label{eq:flux}
\ee
where $F_1$ and $F_2$ are normalization constants and $\delta_i=\alpha+\beta_i$ ($i=1,2$) are the free parameters to be adjusted by fitting to the observed VHE spectrum. Using Eq.(\ref{eq:flux}), we can fit the observed VHE spectra of different EHBL flaring events of Mrk 501. 

The VHE spectra of June 30, 2005 has $0.11\ \mathrm{TeV} \leq E_{\gamma} \leq 6.01\ \mathrm{TeV}$ and we can fit the low energy part of the VHE spectrum (zone-1) with $E^{intd}_{\gamma} \lesssim 1$ TeV very well. The best fit is obtained for $F_1=3.7\times 10^{-10}\ \mathrm{erg\,cm^{-2}\,s^{-1}}$ and $\delta_1=2.77$, with a statistical goodness of 99.9\%. The best fit to the high energy part (zone-2) is obtained for $F_2=3.6\times 10^{-10}\ \mathrm{erg\,cm^{-2}\,s^{-1}}$ and $\delta_2=3.26$, with a statistical significance of 99.7\%. These are shown in Figure \ref{fig:figure1} and compared with the standard photohadronic scenario. It clearly shows that both the zones are distinct and $\delta_1\neq \delta_2$ and these give $\beta_1=0.77$ and $\beta_2=1.26$. Thus we conclude that, the background seed photon fluxes in zone-1 and zone-2 have different behavior.  

\begin{figure}[h]
{\centering
\resizebox*{0.8\textwidth}{0.5\textheight}
{\includegraphics{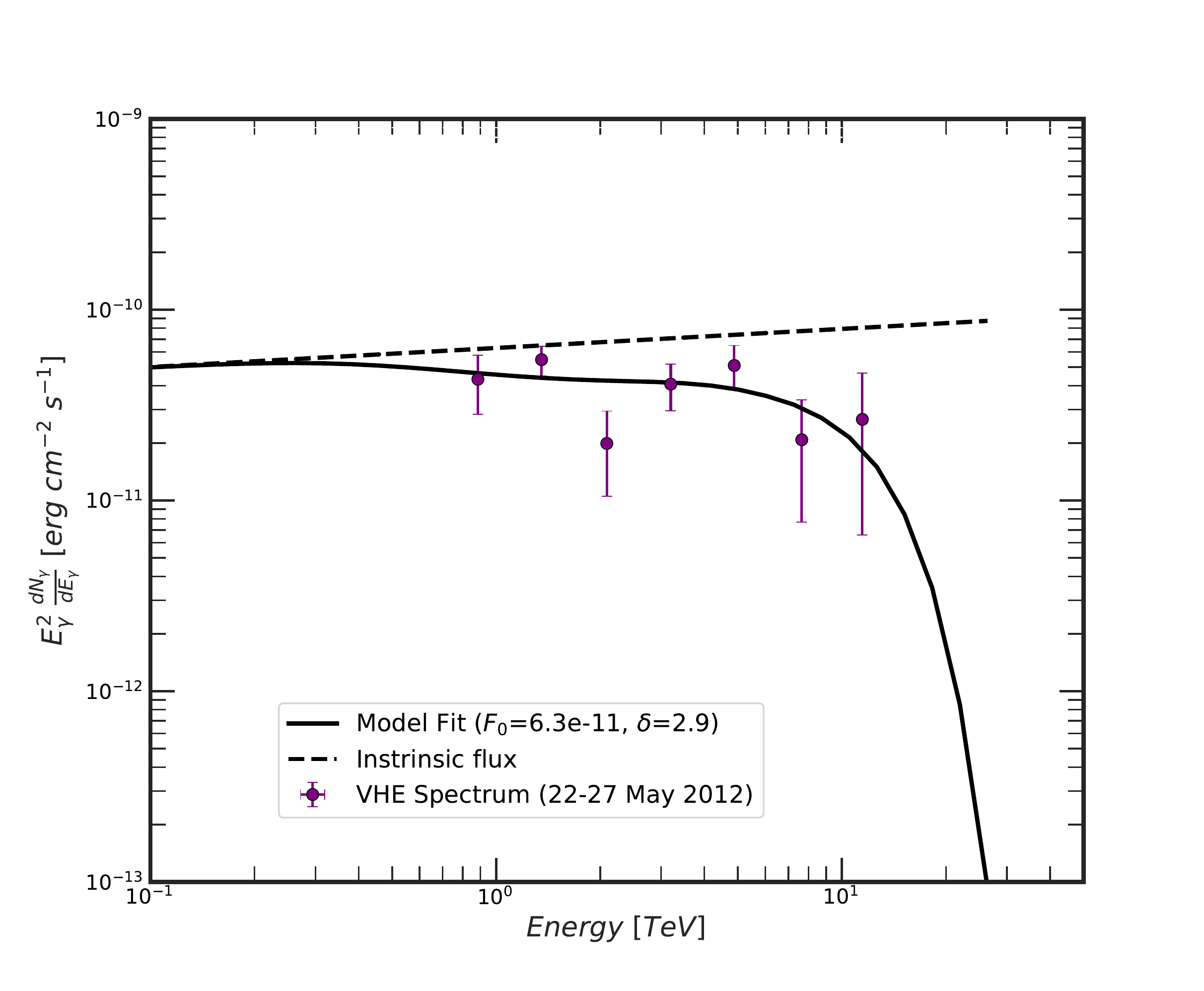}}
\par}
\caption{
The VHE spectrum of Mrk 501 observed during 22-27 May 2012 is fitted with photohadronic model (black curve) and the corresponding intrinsic spectrum is also shown (black dashed). The spectral index $\delta=2.9$ corresponds to high state emission \citep{Sahu:2019kfd}. 
}
\label{fig:figure4}
\end{figure}

\begin{figure}[h]
{\centering
\resizebox*{0.8\textwidth}{0.5\textheight}
{\includegraphics{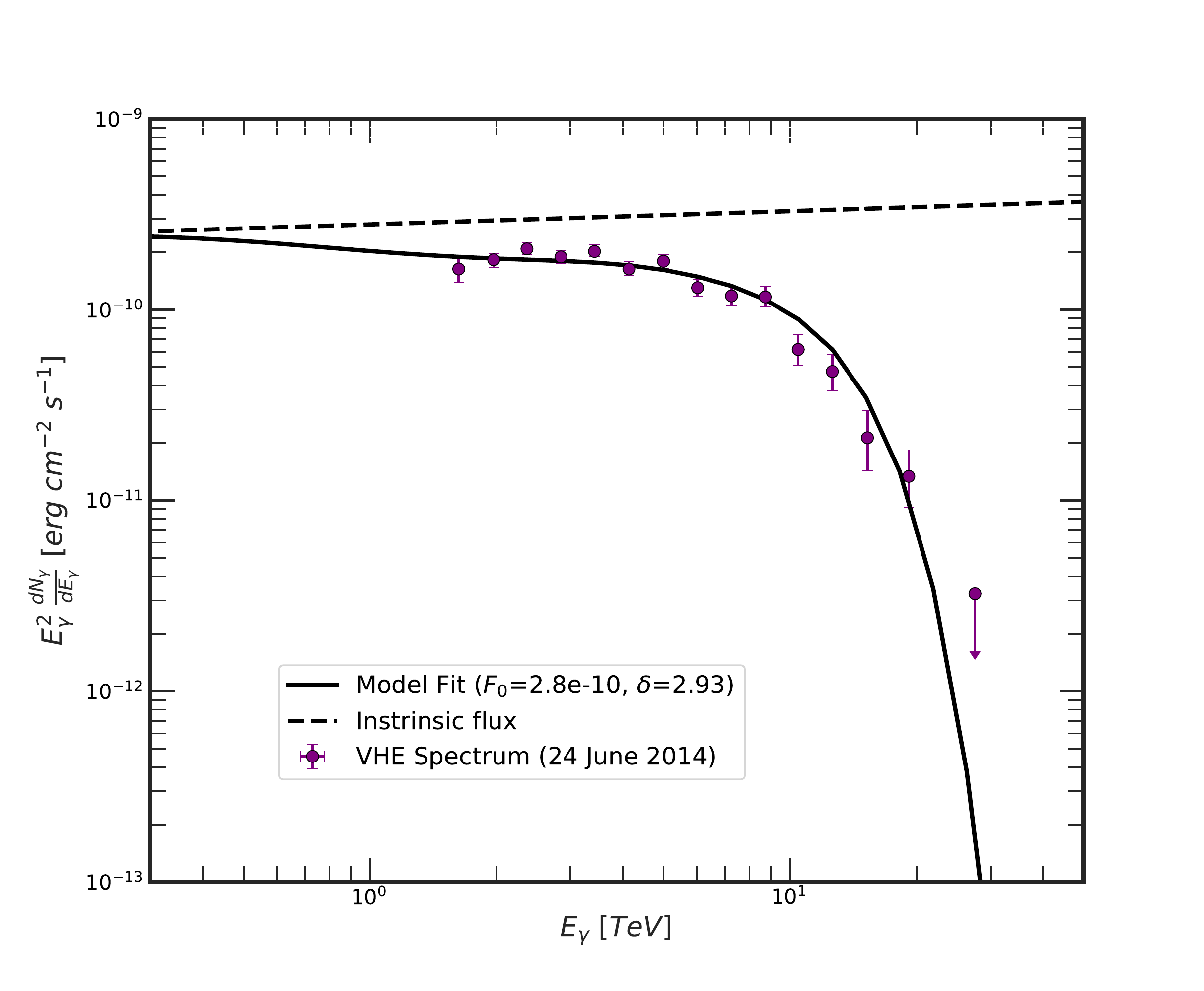}}
\par}
\caption{
The VHE spectrum of Mrk 501 observed in 24 June 2014 is fitted with photohadronic model (black curve) and the corresponding intrinsic spectrum is also shown (black dashed). The spectral index $\delta=2.93$ corresponds to high state emission \citep{Sahu:2019kfd}. 
}
\label{fig:figure5}
\end{figure}

The second VHE flaring was observed on July 9, 2005 in the energy range $0.11\ \mathrm{TeV} \leq E_{\gamma} \leq 4.35\ \mathrm{TeV}$. In zone-1 ($E^{intd}_{\gamma} \lesssim 600$ GeV) the best fit to the spectrum is obtained for $F_1=2.9\times 10^{-10} \mathrm{erg\,cm^{-2}\,s^{-1}}$ and $\delta_1=2.84$, with a statistical goodness of 99.3\%. Similarly in zone-2 the best fit is achieved with
$F_2=2.2\times 10^{-10} \mathrm{erg\,cm^{-2}\,s^{-1}}$ and $\delta_2=3.2$, with a statistical goodness of 99.9\%. Here also the two-zones are distinct, but can be joined smoothly.

Taking two different values of  bulk Lorentz factor ($\Gamma=25$, 50) the multiwavelength SEDs were constructed to explain the flaring events of June 30 and July 9 \citep{Albert:2007zd}. Although these fit the observed spectra well, the minimum values of the SSC energy for both $\Gamma$s are very high ($\epsilon_{\gamma}\simeq 94.6$ MeV and 31.7 MeV respectively). Assuming that these values of  $\epsilon_{\gamma}$ correspond to the maximum observed $E_{\gamma}$, we obtain unrealistically very high values of $\Gamma$ for both the flares (for June 30, $\Gamma\simeq 138,\, 80$ and for July 9, $\Gamma=117,\, 68$). For a reasonable estimate of $\Gamma$, the minimum $\epsilon_{\gamma}$ must be small ($\epsilon_{\gamma} < 10$ MeV).

During the multiwavelength campaign between March and July 2012, the highest activity in VHE was observed on June 9 in the energy range $0.25\ \mathrm{TeV} \leq E_\gamma\leq 6.72\ \mathrm{TeV}$ \citep{Ahnen:2018mtr}. The zone-1 of its VHE spectrum ($E^{intd}_{\gamma} \lesssim 600$ GeV) is fitted very well with  
$F_1=2.9\times 10^{-10}\, \mathrm{erg\,cm^{-2}\,s^{-1}}$ and $\delta_1=2.84$, with a statistical goodness of 99.9\%. Similarly, the best fit to zone-2 is achieved with $F_2=3.7\times 10^{-10} \mathrm{erg\,cm^{-2}\,s^{-1}}$ and $\delta_2=3.5$, with a statistical goodness of 93.8\%. We have also estimated the $\Gamma$ for this flaring event using the multiwavelength SED of one-zone and two-zone leptonic models \citep{Ahnen:2018mtr}. In both cases, the minimum SSC energy is
$\epsilon_{\gamma} \simeq 3.36\times 10^{20}$ Hz ($\sim 1.39$ MeV), which gives $\Gamma\sim 18$, a more reasonable value than previous estimate.
By taking the central black hole mass $M_{\mathrm{BH}}\sim 2\times10^{9}M_{\odot}$, the Eddington luminosity of Mrk 501 is $L_{\mathrm{Edd}}\sim 2.5\times 10^{47}\ \mathrm{erg}\ \mathrm{s^{-1}}$, which has to be equally shared by the jet and the counter-jet. The highest integrated $\gamma$-ray luminosity above 100 GeV is obtained for the June 30, 2005 flaring event, which is $L_{\gamma}\sim 2.4\times 10^{46}\ \mathrm{erg}\ \mathrm{s^{-1}}$. The proton luminosity should satisfy $L_p=7.5\  L_\gamma \tau^{-1}_{p\gamma}< L_{\mathrm{Edd}}/2$, which corresponds to $\tau_{p\gamma}> 0.15$, and by taking $\tau_{p\gamma}\sim 0.2$, we get $L_p\sim 8.9\times 10^{46}\ \mathrm{erg\ s^{-1}}$ which is less than the Eddington luminosity.
 
In the photohadronic model, the maximum proton energy is $E^{\textrm{max}}_p \simeq10\, E^{\textrm{max}}_{\gamma}$. In the above flaring periods we observed that, $E^{\textrm{max}}_{\gamma}\sim 7$ TeV, which corresponds to $E^{\textrm{max}}_p\sim 70$ TeV. Thus, the protons accelerated in the blazar jet are not energetic enough to be observed by cosmic ray detectors on Earth. Additionally, these protons can be deflected by the galactic magnetic field and will be difficult to correlate with the source. Since the EBL energy is very low, in the present context the pion production through photopion process and the subsequent secondary photon production from neutral pions is also negligible.

From the above analysis we observe that, the VHE spectrum in zone-1 is the standard HBL flaring event and all the flaring spectra correspond to high states ($2.6 < \delta_1 < 3.0$). This implies that the seed photon spectral indices for these emission states should have $0.4 < \beta_1 < 1.0$. On the contrary, the spectra in zone-2 have different behavior with $\delta_2 \ge 3.1$. So far 
we have not come across any HBL flaring event that has spectral index $\delta > 3.0$. Thus, $\delta_2 \ge 3.1$ does not correspond to any of the HBL emission states. This is a new  contribution from the EHBL-like behavior of the flaring event. The above value of $\delta_2$ implies $\beta_2 \ge 1.1$. As the proton spectral index is same for both the zones, it is the distribution of the background seed photons in the SSC region which decides the nature of the emission state of a VHE flaring event. Thus, the EHBL nature of the spectrum must be attributed to SSC band where $\beta_2 \ge 1.1$. But, what could be the origin of this higher value of $\beta_2$ ? The leptonic modeling of many HBL SEDs have deep valleys at the junction of  synchrotron spectrum and the SSC spectrum with $\beta\simeq 1.1$ \citep{Sahu:2016mww,Sahu:2018gik}, and Fermi accelerated protons colliding in this region of the seed photon background will produce zone-2 of the spectrum. It seems, the inverse Compton scattering of high energy electrons with the self-produced synchrotron photons during extreme behavior might produce SSC SED with $\beta\ge 1.1$. 

Now question arises, whether, these transitions are temporary or not. To address this, first, we have to analyze some of the flaring events from Mrk 501 before and after these extreme emission states. 
\begin{itemize}
\item Observation of Mrk 501 during a multiwavelength campaign covering a period of 4.5 months from March 15 to August 1, 200 9\citep{Ahnen:2016hsf} by Whipple, VERITAS and MAGIC telescopes observed three different VHE emission states, and all of them are consistent with the standard flaring of an HBL. Also these flaring are best explained using photohadronic model \citep{Sahu:2019scf}.

\item The TACTIC telescope made observation in TeV energy to Mrk 501 between 15 April to 30 May, 2012 for about 70.6 hr and during May 22-27, the source was in high emission state \citep{Chandra:2017vkw}. The observed VHE spectrum was consistent with the HBL emission and can also be explained very well using the photohadronic scenario. We have shown the best fit of our model to the VHE spectrum in Figure \ref{fig:figure4}. However, after a few days, i.e., on 9th of June, the observed VHE emission was EHBL-like.

\item On June 24, 2014, the HESS telescopes observed Mrk 501 for total 1.8 hr in four consecutive runs when rapid flux variability in multi-TeV energy was observed \citep{Abdalla:2019krx}. Once again, this VHE spectrum is perfectly consistent with the HBL flaring and undoubtedly it was in high emission state as explained by photohadronic model \citep{Sahu:2019kfd} which is shown in Figure \ref{fig:figure5}.
\end{itemize}

The above three episodes of standard HBL flaring of Mrk 501 before and after the extreme flaring events of 2005 and 2012 suggest that the EHBL-like events are transient and may repeat in future.

\section{Discussion}

The well known and extensively observed nearby HBL, Mrk 501 has undergone several episodes of multi-TeV flaring since its discovery in 1996. Two flaring events, one in 2005 and another in 2012 are of special significance, since these events were EHBL-like. Usually, the flaring events of HBLs are well explained using the photohadronic model, where the VHE gamma-rays are produced as secondaries from the $p\gamma\rightarrow \Delta^+$ process. However, we found it inadequate to explain the extraneous nature of these EHBL-like events as the VHE flux does not follow a single power-law. In our work, we assume that, even after the transition from HBL to EHBL, the mechanisms of particle acceleration and emission are still the same and photohadronic scenario is the dominant process to produce VHE gamma-rays. We also invoke the two-zones scenario for the SSC seed photons which naturally divides the observed VHE spectrum into two separate zones. Using the two-zones photohadronic model we can explain the VHE spectrum very well. It is observed that, the low energy regime of the spectrum (zone-1) is the standard HBL flaring in VHE with $2.5 < \delta_1 < 3.0$. However, zone-2 has $\delta_2 \ge 3.1$ which is a new contribution due to the EHBL-like nature of the flaring event. We also estimated the bulk Lorentz factor $\Gamma$ of the jet during the extreme flaring epochs which give very high values for the 2005 flaring events and $\Gamma\sim 18$ for the 2012 event.

Many more episodes of multi-TeV flaring from Mrk 501 were observed before and after the EHBL-like events. Undoubtedly all these events are well explained by photohadronic process. From the above observations we argued that EHBL-like events in Mrk 501 are transient and may repeat in future. However, we need to observe this extreme behavior from other well studied HBLs to further corroborate our claim.

We thank Rui Xue and Xiang-Yu Wang for many useful discussions. The work of S.S. is partially supported by
  DGAPA-UNAM (Mexico) Project No. IN103019. S.N. is partially
  supported by ``JSPS Grants-in-Aid for Scientific Research $<$KAKENHI$>$
  (A) 19H00693'', 
``Pioneering Program of RIKEN for Evolution of Matter in the Universe
(r-EMU)'', and 
``Interdisciplinary Theoretical and Mathematical Sciences Program of RIKEN''.

\bibliography{mrk501_ehbl_v0_1}{}

\begin{thebibliography}{}
\expandafter\ifx\csname natexlab\endcsname\relax\def\natexlab#1{#1}\fi
\providecommand{\url}[1]{\href{#1}{#1}}
\providecommand{\dodoi}[1]{doi:~\href{http://doi.org/#1}{\nolinkurl{#1}}}
\providecommand{\doeprint}[1]{\href{http://ascl.net/#1}{\nolinkurl{http://ascl.net/#1}}}
\providecommand{\doarXiv}[1]{\href{https://arxiv.org/abs/#1}{\nolinkurl{https://arxiv.org/abs/#1}}}

\bibitem[{Aartsen {et~al.}(2018{\natexlab{a}})}]{IceCube:2018dnn}
Aartsen, M., {et~al.} 2018{\natexlab{a}}, Science, 361, eaat1378

\bibitem[{Aartsen {et~al.}(2018{\natexlab{b}})}]{IceCube:2018cha}
---. 2018{\natexlab{b}}, Science, 361, 147

\bibitem[{Abdalla {et~al.}(2019)}]{Abdalla:2019krx}
Abdalla, H., {et~al.} 2019, Astrophys. J., 870, 93

\bibitem[{Abdo(2010)}]{Abdo:2010rw}
Abdo, A. 2010, Astrophys. J., 722, 520

\bibitem[{Abdo {et~al.}(2010)}]{Abdo:2009iq}
Abdo, A., {et~al.} 2010, Astrophys. J., 716, 30

\bibitem[{Abdo {et~al.}(2009)}]{Abdo:2009wu}
Abdo, A.~A., {et~al.} 2009, Astrophys. J., 700, 597

\bibitem[{Acciari {et~al.}(2011)}]{Acciari:2010aa}
Acciari, V., {et~al.} 2011, Astrophys. J., 729, 2

\bibitem[{Acciari {et~al.}(2019)}]{Acciari:2019ntl}
---. 2019, Mon. Not. Roy. Astron. Soc., 490, 2284

\bibitem[{Acciari {et~al.}(2020)}]{Hayashida:2020wez}
---. 2020, Astron. Astrophys., 638, A14

\bibitem[{Ackermann {et~al.}(2012)}]{Ackermann:2012sza}
Ackermann, M., {et~al.} 2012, Science, 338, 1190

\bibitem[{Ahnen {et~al.}(2017)}]{Ahnen:2016hsf}
Ahnen, M., {et~al.} 2017, Astron. Astrophys., 603, A31

\bibitem[{Ahnen {et~al.}(2018)}]{Ahnen:2018mtr}
---. 2018, Astron. Astrophys., 620, A181

\bibitem[{Albert {et~al.}(2007)}]{Albert:2007zd}
Albert, J., {et~al.} 2007, Astrophys. J., 669, 862

\bibitem[{Albert {et~al.}(2008)}]{Albert:2007me}
---. 2008, Astrophys. J., 681, 944

\bibitem[{Aleksi\'c {et~al.}(2015)}]{Aleksic:2014usa}
Aleksi\'c, J., {et~al.} 2015, Astron. Astrophys., 573, A50

\bibitem[{Aliu {et~al.}(2016)}]{Aliu:2016kzx}
Aliu, E., {et~al.} 2016, Astron. Astrophys., 594, A76

\bibitem[{Blazejowski {et~al.}(2000)Blazejowski, Sikora, Moderski, \&
  Madejski}]{Blazejowski:2000ck}
Blazejowski, M., Sikora, M., Moderski, R., \& Madejski, G. 2000, Astrophys. J.,
  545, 107

\bibitem[{Böttcher {et~al.}(2008)Böttcher, Dermer, \&
  Finke}]{Boettcher:2008fh}
Böttcher, M., Dermer, C.~D., \& Finke, J.~D. 2008, Astrophys. J. Lett., 679,
  L9

\bibitem[{Cao \& Wang(2014)}]{Cao:2014nia}
Cao, G., \& Wang, J. 2014, Astrophys. J., 783, 108

\bibitem[{Chandra {et~al.}(2017)}]{Chandra:2017vkw}
Chandra, P., {et~al.} 2017, New Astron., 54, 42

\bibitem[{Costamante {et~al.}(2018)Costamante, Bonnoli, Tavecchio, Ghisellini,
  Tagliaferri, \& Khangulyan}]{Costamante:2017xqg}
Costamante, L., Bonnoli, G., Tavecchio, F., {et~al.} 2018, Mon. Not. Roy.
  Astron. Soc., 477, 4257

\bibitem[{Costamante {et~al.}(2001)}]{Costamante:2001pu}
Costamante, L., {et~al.} 2001, Astron. Astrophys., 371, 512

\bibitem[{Dermer \& Schlickeiser(1993)}]{Dermer:1993cz}
Dermer, C.~D., \& Schlickeiser, R. 1993, Astrophys. J., 416, 458

\bibitem[{Dominguez {et~al.}(2011)}]{Dominguez:2010bv}
Dominguez, A., {et~al.} 2011, Mon. Not. Roy. Astron. Soc., 410, 2556

\bibitem[{Foffano {et~al.}(2019)Foffano, Prandini, Franceschini, \&
  Paiano}]{Foffano:2019itc}
Foffano, L., Prandini, E., Franceschini, A., \& Paiano, S. 2019, Mon. Not. Roy.
  Astron. Soc., 486, 1741

\bibitem[{Franceschini {et~al.}(2008)Franceschini, Rodighiero, \&
  Vaccari}]{Franceschini:2008tp}
Franceschini, A., Rodighiero, G., \& Vaccari, M. 2008, Astron. Astrophys., 487,
  837

\bibitem[{Gao {et~al.}(2013)Gao, Lei, \& Zhang}]{Gao:2012sq}
Gao, H., Lei, W.-H., \& Zhang, B. 2013, Mon. Not. Roy. Astron. Soc., 435, 2520

\bibitem[{Ghisellini {et~al.}(2005)Ghisellini, Tavecchio, \&
  Chiaberge}]{Ghisellini:2004ec}
Ghisellini, G., Tavecchio, F., \& Chiaberge, M. 2005, Astron. Astrophys., 432,
  401

\bibitem[{Kataoka {et~al.}(1999)}]{Kataoka:1998mw}
Kataoka, J., {et~al.} 1999, Astrophys. J., 514, 138

\bibitem[{Kranich(2009)}]{Kranich:2009ht}
Kranich, D. 2009.
\newblock \doarXiv{0912.3830}

\bibitem[{Maraschi {et~al.}(1992)Maraschi, Ghisellini, \&
  Celotti}]{Maraschi:1992iz}
Maraschi, L., Ghisellini, G., \& Celotti, A. 1992, Astrophys. J. Lett., 397, L5

\bibitem[{Murase {et~al.}(2012)Murase, Dermer, Takami, \&
  Migliori}]{Murase:2011cy}
Murase, K., Dermer, C.~D., Takami, H., \& Migliori, G. 2012, Astrophys. J.,
  749, 63

\bibitem[{Padovani {et~al.}(2016)Padovani, Resconi, Giommi, Arsioli, \&
  Chang}]{Padovani:2016wwn}
Padovani, P., Resconi, E., Giommi, P., Arsioli, B., \& Chang, Y. 2016, Mon.
  Not. Roy. Astron. Soc., 457, 3582

\bibitem[{Paggi {et~al.}(2009)Paggi, Massaro, Vittorini, Cavaliere, D'Ammando,
  Vagnetti, \& Tavani}]{Paggi:2009yx}
Paggi, A., Massaro, F., Vittorini, V., {et~al.} 2009, Astron. Astrophys., 504,
  821

\bibitem[{Petry {et~al.}(2000)}]{Petry:2000pb}
Petry, D., {et~al.} 2000, Astrophys. J., 536, 742

\bibitem[{Sahu(2019)}]{Sahu:2019lwj}
Sahu, S. 2019, Rev. Mex. Fis., 65, 307

\bibitem[{Sahu {et~al.}(2018)Sahu, de~León, Nagataki, \& Gupta}]{Sahu:2018gik}
Sahu, S., de~León, A.~R., Nagataki, S., \& Gupta, V. 2018, Eur. Phys. J. C,
  78, 557

\bibitem[{Sahu {et~al.}(2020)Sahu, López~Fortín, Iglesias~Martínez,
  Nagataki, Fernández~de Córdoba, Iglesias~Martínez, \& Fernández~de
  Córdoba}]{Sahu:2019scf}
Sahu, S., López~Fortín, C., Iglesias~Martínez, M., {et~al.} 2020, Mon. Not.
  Roy. Astron. Soc., 492, 2261

\bibitem[{Sahu {et~al.}(2019)Sahu, López~Fortín, \& Nagataki}]{Sahu:2019kfd}
Sahu, S., López~Fortín, C.~E., \& Nagataki, S. 2019, Astrophys. J. Lett.,
  884, L17

\bibitem[{Sahu {et~al.}(2017)Sahu, Yáñez, Miranda, de~León, \&
  Gupta}]{Sahu:2016mww}
Sahu, S., Yáñez, M. V.~L., Miranda, L.~S., de~León, A.~R., \& Gupta, V.
  2017, Eur. Phys. J. C, 77, 18

\bibitem[{Sikora {et~al.}(1994)Sikora, Begelman, \& Rees}]{Sikora:1994zb}
Sikora, M., Begelman, M.~C., \& Rees, M.~J. 1994, Astrophys. J., 421, 153

\bibitem[{Tavecchio(2014)}]{Tavecchio:2013fwa}
Tavecchio, F. 2014, Mon. Not. Roy. Astron. Soc., 438, 3255

\bibitem[{Tavecchio \& Bonnoli(2015)}]{Tavecchio:2015cid}
Tavecchio, F., \& Bonnoli, G. 2015.
\newblock \doarXiv{1512.05080}

\bibitem[{Urry \& Padovani(1995)}]{Urry:1995mg}
Urry, C., \& Padovani, P. 1995, Publ. Astron. Soc. Pac., 107, 803

\end{thebibliography}
\bibliographystyle{aasjournal}

\end{document}